\documentclass[aip,preprint]{revtex4-1}
\usepackage{graphicx}
\usepackage{caption}
\usepackage{dcolumn}
\usepackage{bm}
\usepackage{ulem}

\begin{document}

\preprint{AIP/123-QED}

\title{Nonlinear evolution of electron shear flow instabilities in the presence of an external guide magnetic field}

\author{Neeraj Jain}
\affiliation{ Max Planck Institute for Solar System Research,
Justus-von-Liebig-Weg 3, 37077, G\"ottingen, Germany.}
\author{J\"org B\"uchner}%
\affiliation{ Max Planck Institute for Solar System Research,
Justus-von-Liebig-Weg 3, 37077, G\"ottingen, Germany.}

\date{\today}

\begin{abstract}
The dissipation mechanism by which the magnetic field reconnects  in the presence of an external (guide) magnetic field in the direction of the main current is not well understood.
 In thin electron current sheets (half thickness close to an electron inertial length) formed in collisionless magnetic reconnection, electron shear flow instabilities are potential candidates for providing an anomalous dissipation mechanism which can break the frozen-in condition of the magnetic field affecting the structure and rate of reconnection. 
We present the results of investigations of the evolution of electron shear flow instabilities, from linear to nonlinear state, in guide field magnetic reconnection. The properties of the plasma turbulence resulting from the growth of the instability and their dependence on the strength of the guide field are studied. For this sake, we utilize three dimensional electron-magnetohydrodynamic simulations  of electron current sheets.   We show that, unlike the case of current sheets self-consistently embedded in anti-parallel magnetic fields,  the evolution of thin electron current sheets in the presence of a  finite external guide field (equal to the asymptotic value of the reconnecting magnetic field or larger) is dominated by high wave number non-tearing mode instabilities. The latter cause the development of, first, a wavy structure of the current sheet. The turbulence, developed later, consists of  current filaments and electron flow vortices.  As a result of the nonlinear evolution of the instability, the current sheet broadens simultaneously with its  flattening in the central region mimicking a  viscous-like turbulent dissipation. Later, the flattened current sheet bifurcates. During the time of the bifurcation, the rate of the change of the mean electron flow velocity is proportional to the magnitude of the flow velocity, suggesting a resistive-like dissipation.
The turbulence energy cascades to shorter wavelengths preferentially in the direction perpendicular to the guide magnetic field. The degree of anisotropy of the turbulence was found to increase  with the increasing strength of the guide field.  
\end{abstract}

\pacs{Valid PACS appear here}
\maketitle

\section{\label{sec:introduction}Introduction}
In a variety of plasma situations, e.g., in collisionless magnetic reconnection, fast Z-pinches, plasma  opening switches etc. \cite{birnbook,ryutov2000,kingsep90}, thin current sheets due to electron shear flows develop with a characteristic transverse length scale  of the order of few electron inertial lengths.
In collision-less magnetic reconnection, e.g., the dissipation region in which magnetic fluxes reconnect become very thin  and develops  a two scale structure \cite{hesse99}. In its inner part, an electron
current sheet (ECS) with a thickness of the order of an electron inertial length
($d_e=c/\omega_{pe}$) forms. The ECS is embedded inside an ion current sheet  with thickness of the
order of an ion inertial length, $d_i=c/\omega_{pi}$ \cite{hesse99}. The topology change in the magnetic field required for magnetic reconnection, however, takes place inside the embedded thin electron current sheets in which the electrons are the dominant carriers of the current. 

In case of anti-parallel magnetic reconnection, non-gyrotropic orbits of electrons caused by their demagnetization in the field reversal region provide the necessary dissipation for the topology changes of the magnetic field. In case of guide (external) magnetic field reconnection, however, electrons remain magnetized throughout the whole current sheet and the nature of the dissipation must be different. The micro-instabilities of electron current sheets may allow magnetic reconnection  by generating electro-magnetic turbulence, and thus, anomalous dissipation which in turn interacts with the electrons breaking their flows and gyrotropy. Spacecraft observations, e.g., in the Earth's magnetotail and laboratory experiments  of magnetic reconnection have revealed the existence of electron shear flows in these electron-scale wide inner current sheets \cite{wygant05,dorfman2013,jain2013,stechow2015}. Three dimensional PIC simulations of magnetic reconnection, on the other hand, have revealed the electron shear flow instabilities in the current sheets \cite{che2011}.



In particular, electron shear flows can be unstable to
3-D oblique tearing
modes which directly generate flux ropes/plasmoids \cite{daughton2011,markidis2013}
and/or to non-tearing modes which filament the ECS \cite{che2011}. In the presence of an external (guide) magnetic field in the direction of the electron flow, the
conditions under
which an ECS filaments and/or generates flux ropes/plasmoids depend on the magnetic-guide-field strength and on the thickness of the current sheet. As we have shown earlier in the limit of zero guide field (anti-parallel magnetic field), the tearing mode instability dominates over non-tearing modes only if the ECS are thin enough (with a half thickness 
close to an electron
inertial length) \cite{jain2014a,jain2014b}. Otherwise non-tearing mode instabilities dominate. In recent eigen-value analysis of 3-D electron shear flow instabilities, we have shown that finite guide field makes the  linear electron tearing mode instability sub-dominant even for the thin current sheets \cite{jain2015}. 

In this paper, we now report the results of investigations  of the  nonlinear evolution of electron shear flow instabilities in ECS in the presence of and dependence on finite guide magnetic field. Using 3-D electron-magnetohydrodynamic\cite{kingsep90} simulations, we found that the evolution is dominated by non-tearing modes for thin current sheets with half-thickness equal to an electron inertial length and a finite guide field.  In the course of the nonlinear evolution, the current sheet instability develops turbulence consisting of electron flow vortices and current filaments. The developed turbulence is anisotropic with respect to the direction of the guide field. It provides resistive and viscous like dissipation to the electron flow.
     
Investigations of  linear and
nonlinear electron shear flow instabilities using an EMHD approximation
have been reported earlier
\cite{basova1991,das01,drake94,jain03,jain04,jain2012b,gaur2012}. In
EMHD the electron
flow carries the current. Hence EMHD instabilities are also referred  to as current driven (sausage and/or kink depending upon the wave mode structures) instabilities \cite{das01,jain03,jain04}. The growth
rates of 2-D sausage (in the plane containing
directions of flow and shear) and kink (in the plane perpendicular to the
electron flow) modes 
were obtained \cite{gaur2012}. 
Most of these studies, however, focus on equilibrium configurations in
which none of the
magnetic
field component changes its sign across current sheets and where the current sheet
thickness is  much less than an electron inertial length.  In contrast to
these investigations,
our goal in this paper is to study the nonlinear 
3-D evolution of the instabilities of ECSs with thicknesses $\geq d_e$ in a finite-guide-field current-sheet configuration suitable for component magnetic reconnection.

In the next section we introduce the electron-magnetohydrodynamic (EMHD) model used for our analysis and describe the simulation setup. The evolution of the electron shear flow instabilities in the presence of a finite guide field, from linearly unstable phase to nonlinear saturation,  is presented in  section \ref{sec:evolution}. We summarize our findings  in
section \ref{sec:conclusion}.

\section{\label{sec:emhd}Electron-MHD model and simulation setup}
We use an electron-magnetohydrodynamic  (EMHD) model \cite{kingsep90} which considers the electron dynamics  in a stationary ion background in fluid approximation. Such approximation of the plasma dynamics  is valid
for processes at spatial scales smaller than $d_i$ and time scales smaller than
$\omega_{ci}^{-1}$. It solves the electron momentum equation coupled with Maxwell's equations. The evolution equation for the magnetic field $\mathbf{B}$ is obtained by eliminating 
the electric field from the electron momentum equation using Faraday's law \cite{kingsep90}.
\begin{eqnarray}
\frac{\partial}{\partial t}(\mathbf{B}-d_e^2\nabla^2\mathbf{B})&=&\nabla \times
[\mathbf{v}_e\times (\mathbf{B}-d_e^2\nabla^2\mathbf{B})]\label{eq:emhd1},
\end{eqnarray}
where, $\mathbf{v}_e=-(\nabla\times\mathbf{B})/\mu_0n_0e$ is the electron fluid velocity. In addition to
ignoring the ion dynamics, Eq. (\ref{eq:emhd1}) assumes a uniform electron
number
density $n_0$ and the incompressibility of the electron fluid. Assuming $\omega << \omega_{pe}^2/\omega_{ce}$, displacement
currents are ignored.
In EMHD, the frozen-in condition of magnetic fluxes breaks down due to the
finite electron inertia (which is contained in the definition of $d_e \propto
\sqrt{m_e}$). In the absence of electron
inertia ($d_e\rightarrow 0$), Eq. (\ref{eq:emhd1}) represents the frozen-in condition of magnetic flux in an ideal electron flow. 

The equilibrium magnetic field is chosen to be
\begin{eqnarray}
\mathbf{B}=B_{0}\tanh(x/L)\hat{y}+B_{g}\hat{z}
\end{eqnarray}
corresponding to a current density
$\mathbf{J}_{eq}=(B_0/\mu_0L)\mathrm{sech}^2(x/L)\hat{z}$ where $B_0$ is the asymptotic value ($x>>L$) of the reconnecting component (y) of the magnetic field and $B_g$ is the external guide magnetic field in the z-direction. The y-component of the equilibrium magnetic field reverses its direction across the  ECS with a half thickness $L$. 
Since ions are considered as stationary background,
the electron fluid velocity is related to the current density by the relation  $\mathbf{J}=-n_0e\mathbf{v}_{e}$.

In the simulations, an initial small-amplitude perturbation  is added to the equilibrium to hasten  the instabilities. It consists of a spectrum of wave numbers given by
\begin{eqnarray}
\tilde{f}(x,y,z,t=0)&=&f_1\exp\left(-\frac{x^2}{L^2}\right)
\sum_{m,n=1}^{m,n=20}
\sin\left(\frac{m\pi y}{L_y}\right)
\sin\left(\frac{n\pi z}{L_z}\right)
\end{eqnarray}
where $f_1$ is the amplitude of the initial perturbation, and, 
$L_y$ and $L_z$ are the half-lengths of the simulation box in the y- and z-directions, respectively. 
The initial  perturbation $\tilde{f}$ inputs a finite power into $20\times20$ wave modes.
It is chosen to be maximum at the center of the
simulation domain which extends from $x = -L_x$ to $L_x$, $y = -L_y$ to $L_y$ and
$z =- L_z$ to $L_z$. The boundary conditions are chosen to be periodic along the y and z directions. The perturbations vanish at the remote boundaries in the x-direction far away from the central region of
interest. 

The current sheet half-thickness is taken to be equal to an electron inertial length, $L=d_e$. The strength of the guide field is varied in the range  $B_g/B_0$=1-4. The length of the simulation box and the grid spacing along the x-direction are $2\,L_x=30\,L$ and $\Delta_x=L/5$, respectively. 
Along the y- and z-directions, the lengths of the simulation box, $2L_{y,z}$, and the grid spacing, $\Delta_{y,z}$,  are chosen as $2\,\pi/k_{y,z}^{min}$ and $\pi/k_{y,z}^{max}$, respectively, in order to include a sufficiently large number of  the potentially unstable tearing and non-tearing modes. The upper ($k_y^{max}$ and $k_z^{max}$) and lower ($k_y^{min}$ and $k_z^{min}$) bounds on $k_y$ and $k_z$ of the unstable modes  were obtained by carrying out a normal eigen-mode analysis of the linearized EMHD equations for each of the given values of the $L$ and $B_g$ \cite{jain2015}. Initially, the time step $\Delta_t$ is set equal to the maximum possible value $\Delta_t^{max}$ still allowed in a simulation for given values of $L$ and $B_g$, i.e., 
\begin{eqnarray}
\Delta_t^{max}&=& \mathrm{MIN}
\left[
\frac{0.05}{f_{wh}^{max}},
\frac{0.05}{\gamma_{max}},
\frac{\mathrm{MIN}(\Delta_x,\Delta_y,\Delta_z)\,C}{v_{max}}
\right]
\label{eq:dtmax}
\end{eqnarray}
In expression (\ref{eq:dtmax}), $f_{wh}^{max}=eB_{\infty}/2\pi m_e$ is the maximum whistler frequency in the asymptotic magnetic field $B_{\infty}=(B_0^2+B_g^2)^{1/2}$, $v_{max}=\mathrm{MAX}[v_{wh}^{max},v_{flow}^{max}]$, $v_{wh}^{max}=B_{\infty}/2\sqrt{\mu_0n_0m_e}$ is the maximum phase velocity of whistlers in the asymptotic magnetic field $B_{\infty}$, $v_{flow}^{max}=d_e^2eB_0/m_eL$ (maximum electron flow velocity), $\gamma_{max}$ is the maximum growth rate of the instability and $C$ (=0.5) is the Courant-condition number. A smaller time step is used during the simulations if required by the Courant condition corresponding to  the maximum  instantaneous electron flow velocity (third term on the RHS of expression (\ref{eq:dtmax}) ). The simulations are run up to the time $T_f=40/\gamma_{max}$ which is still small enough for neglecting the ion dynamics but large enough to fully follow the nonlinear evolution of the electron shear flow instabilities till their saturation. 

 Results are presented in normalized units: the magnetic field
 is normalized by $B_0$ , the length by the electron skin depth
$d_e$, the time by the inverse electron cyclotron frequency
$\omega_{ce}^{-1}=(eB_0/m_e)^{-1}$ , hence, the velocity by $v_{Ae}=d_e\omega_{ce}$, the electron Alfv\'en
velocity. Note that under this normalization
$\mathbf{J}=-\mathbf{v}_e$ .


\section{Evolution of electron shear flow instabilities in the presence of a guide field \label{sec:evolution}}
\begin{figure}
\includegraphics[width=0.5\textwidth,height=0.3\textheight]{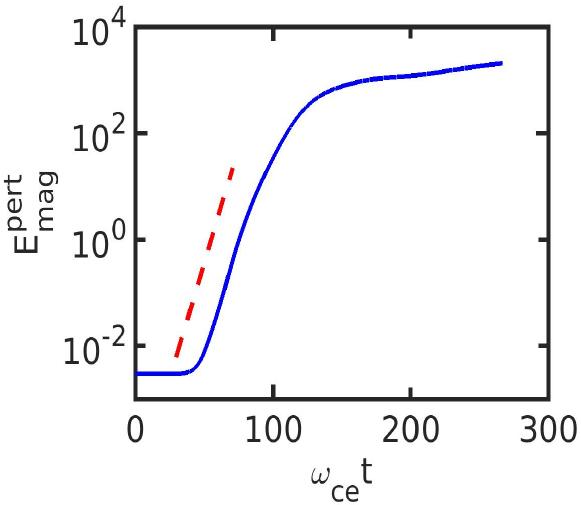}


\caption{\label{fig:linear_phase} 
Evolution of volume integrated energy in the magnetic field perturbations for $L=d_e$ and $B_g=B_0$ (solid line). The slope of the dashed line  is twice the maximum growth rate, obtained from the linear theory, of the initialized modes. 
}
\end{figure}

We present detailed results of the investigations of the 3-D evolution of electron shear flow instabilities from linear to non-linearly saturated state for a representative case of $L=d_e$ and $B_g=B_0$. 
Fig. \ref{fig:linear_phase} shows the evolution of the volume integrated  perturbed magnetic field energy, $E_{mag}^{pert}=\int (\tilde{B}_x^2+\tilde{B}_y^2+\tilde{B}_z^2)\, dx\, dy\, dz$. 
After an initial transient phase, i.e. after $\omega_{ce}t \approx 40$, the energy of the perturbations grows exponentially with time, $E_{mag}^{pert}\propto e^{2\gamma t}$, where $\gamma$ is the maximum growth rate of the initialized modes. The slope of the dashed line ($2\gamma$) is in good agreement with slope of the energy evolution in Fig. \ref{fig:linear_phase}. The growth of the perturbed energy saturates after about $\omega_{ce}t=120$.

\begin{figure}
\includegraphics[width=0.7\textwidth,height=0.8\textheight]{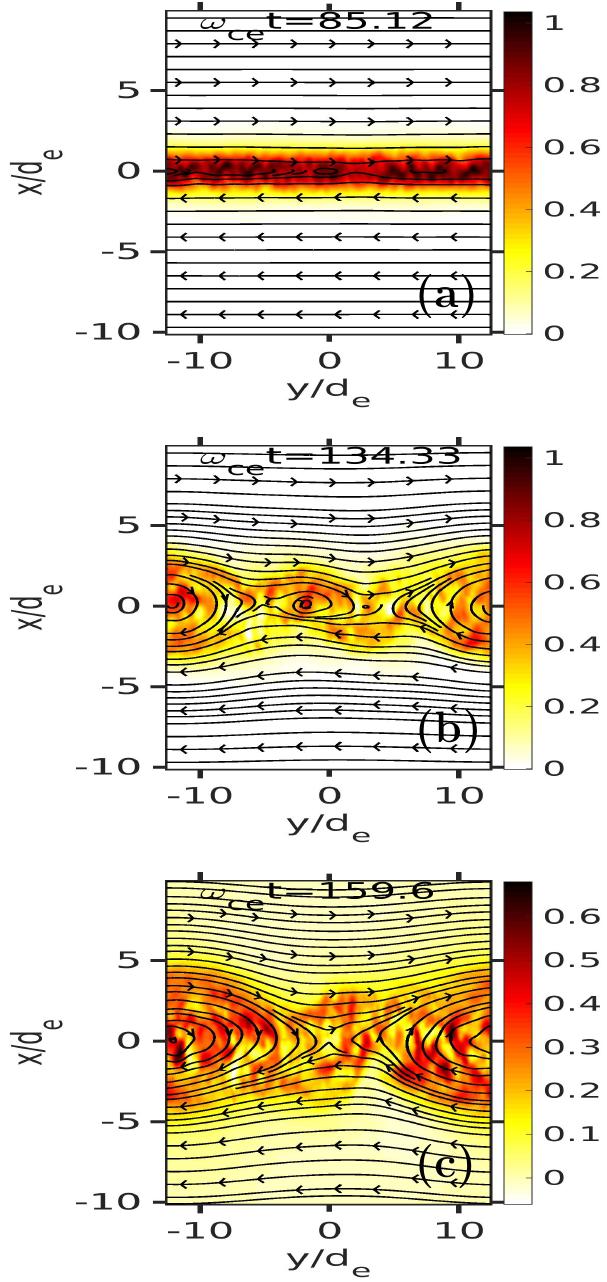}


\caption{\label{fig:jz_xy} Case of  a thin current sheet ($L=d_e$) and a guide field $B_g=B_0$. Current density $J_z$ (color coded, legend to the right) in x-y plane (z=0) at three different times.  A projection of the magnetic field lines into the x-y reconnection plane is over-plotted.
}
\end{figure}

\begin{figure}
\includegraphics[width=0.7\textwidth,height=0.8\textheight]{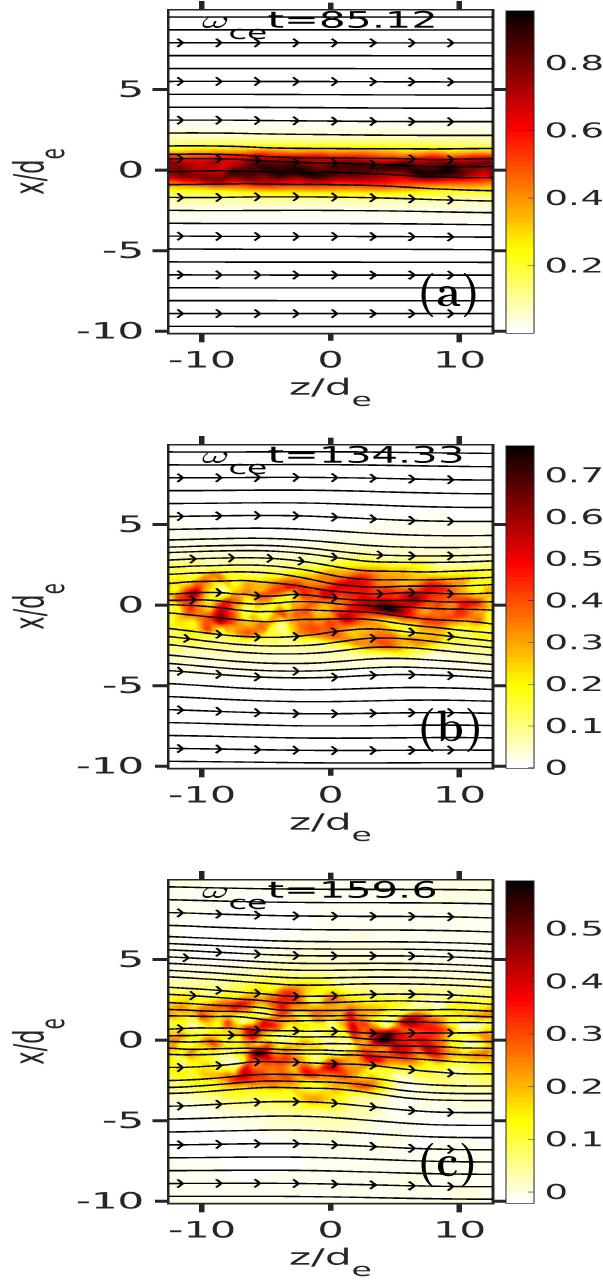}



\caption{\label{fig:jz_xz} Current density $J_z$ (color) in x-z plane (y=0) at three different times for $L=d_e$ and $B_g=B_0$. A projection of the magnetic field lines into the x-z plane is over-plotted.
}
\end{figure}

Figs. \ref{fig:jz_xy} and \ref{fig:jz_xz} show the evolution of current density $J_z$ in the x-y and x-z planes.
The current sheet develops short-wavelength structures along the y-direction starting at $\omega_{ce}t \approx 85$. The wavelength along the guide field direction (z) is larger than that developed along the y-direction, ($k_z < k_y$). This is due to the growth of the non-tearing mode with a large $k_y$ and relatively small but finite $k_z$. While at later times an unstable tearing mode also develops, the ECS evolution has already been  dominated by the growth of non-tearing modes which first cause the wavy structure of the ECS (see Fig. \ref{fig:jz_xy}a) and later turbulence consisting of current filaments and electron vortices. The tearing mode develops in the turbulent ECS. Note that this is in strong contrast to the evolution of thin current sheet ($L=d_e$) in anti-parallel magnetic fields in which the tearing mode instability develops in a laminar ECS and dominates the evolution \cite{jain2014b}. On the other hand, the ECS evolution in case of a finite guide field reminds that of  thick current sheets with vanishing guide fields \cite{jain2014b}. 

\begin{figure}
\includegraphics[width=0.7\textwidth,height=0.8\textheight]{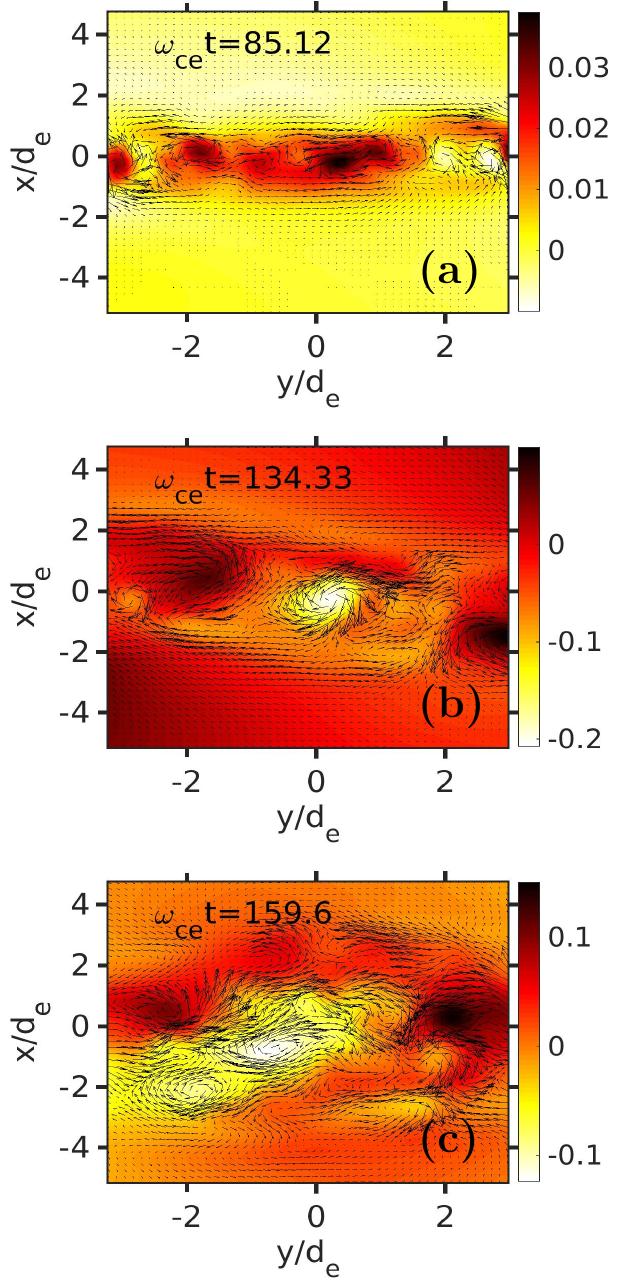}


\caption{\label{fig:bz} The magnetic field component $B_z$ (color) and electron flow  vectors in x-y plane (z=0) at three different times for $L=d_e$ and $B_g=B_0$.
}
\end{figure}

\begin{figure}
\includegraphics[width=0.7\textwidth,height=0.8\textheight]{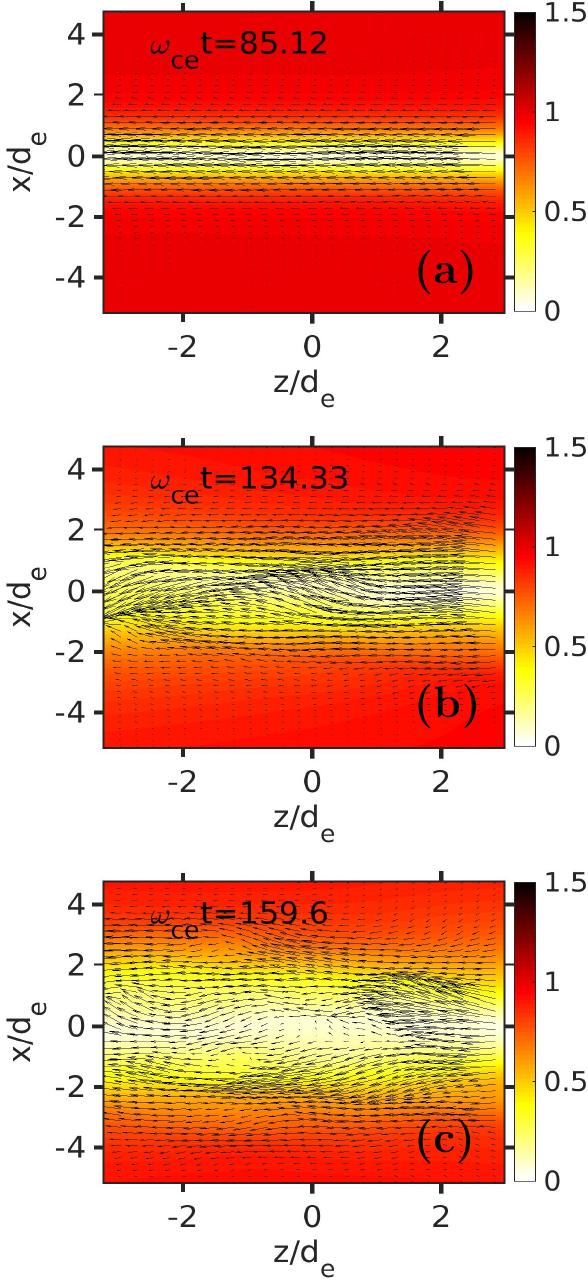}


\caption{\label{fig:b} Magnitude of the total magnetic field (color) and electron flow vectors in the x-z plane (y=0) at three different times  for $L=d_e$ and $B_g=B_0$.
}
\end{figure}

The development of electron vortices is illustrated in Fig. \ref{fig:bz} which shows the magnetic field component $B_z$ and electron flow vectors in the central region of the x-y ("reconnection'') plane (z=0). At $\omega_{ce}t\approx 85$, the electron flow vectors $v_x$ and $v_y$ develop vortices in the bulk current flow. Then the vortices interact with each other and grow in size. In addition after $\omega_{ce}t\approx 160$ new electron vortices develop at the edges of the current sheet.  The nonlinear evolution of the linearly unstable non-tearing modes  results, finally, in a broadening of the current sheet. Fig. \ref{fig:b} depicts the magnitude of the total magnetic field $|B|=(B_x^2+B_y^2+B_z^2)^{1/2}$ and the electron flow vectors in x-z plane. The velocity component $v_x$ did not grow to an appreciable value (in comparison to $v_z$) by $\omega_{ce}t=85$ and thus the electron flow is mainly directed along z-direction. At later times, electrons start to flow along the curved paths in the x-z plane. As a result, z-directed filaments of the electron current forms.

\begin{figure}
\includegraphics[width=0.7\textwidth,height=0.8\textheight]{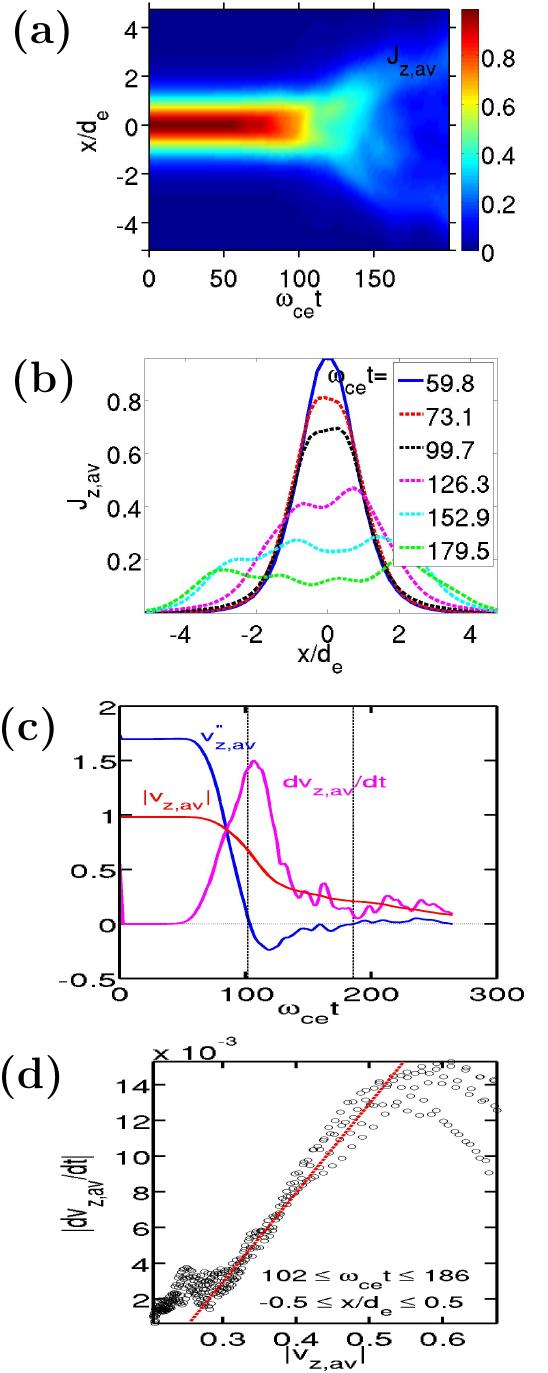}



\caption{\label{fig:jzav} Temporal evolution of the average current density distribution $J_{z,av}$ obtained by averaging over the y- and z-directions, shown in x-t plane (a) and its line-outs along x at different times (b) for $L=d_e$ and $B_g=B_0$. Temporal evolution of $v_{z,av}^{''}=d^2v_{z,av}/dx^2$, $|v_{z,av}|$ and 100$\times dv_{z,av}/dt$ at x=0 (c). Scatter plot of $|dv_{z,av}/dt|$ vs $|v_{z,av}|$ at the locations $-0.5 \leq x/d_e \leq 0.5$ during the time interval $102 \leq \omega_{ce}t \leq 186$ (d).
}
\end{figure}

In order to assess the role of the turbulence generated by the electron shear flow instabilities in providing an anomalous dissipation, we show in Fig. \ref{fig:jzav}a the temporal evolution of $J_{z,av}=(1/4L_yL_z)\int J_z dy dz$, the current density averaged along the y- and z- directions. As one can see in the figure, the averaged current density begins to drop in the central region at around $\omega_{ce}t=60$ while its broadening starts  after $\omega_{ce}t\approx 100$. Finally the current density $J_{z,av}$ drops significantly to small values in the central region. At the edges, $J_{z,av}$ is slightly enhanced  compared to its value in the central region. In Fig. \ref{fig:jzav}b we show the temporal evolution  of the x-profiles of $J_{z,av}$. Until $\omega_{ce}t\approx 100$, $J_{z,av}$ drops and flattens  mainly in the peak region of the average current profile where its curvature is maximum. Once the current profile is curvature-free in the central region, $J_{z,av}$ at $x=0$ drops faster than at the neighboring points, bifurcating the average current profile. Bifurcated electron current sheets have also been observed in the 3-D particle-in-cell simulations of force free electron current sheets \cite{liu2013}. The bifurcation in these PIC simulations, however, was attributed to the development of the tearing mode at multiple resonance surfaces. Simultaneous with the bifurcation in the central region,   $J_{z,av}$ enhances at the edges where (again) the second derivative i.e. the curvature of the current profile is large. The enhancement of the current $J_{z,av}$ at the edges broadens the current sheet. 

A broadening at the edges and a flattening in central (maximum current) region is typical for  a viscous dissipation. However the localized drop of the flattened $J_{z,av}$ in the vicinity of $x=0$, bifurcating the current sheet there, requires a dissipation mechanism which can operate without the need of the curvature of the current density profile. In order to understand the dissipation mechanism responsible for the bifurcation, we show in Fig. \ref{fig:jzav}c the evolution of the absolute value of $v_{z,av}=-J_{z,av}$,  its second order spatial derivative $v_{z,av}^{''}=d^2v_{z,av}/dx^2$ (a measure of the curvature) and the time derivative $dv_{z,av}/dt$ at  $x\approx 0$. The rate of the current drop $dv_{z,av}/dt$ decreases during the current sheet bifurcation, i.e., when $v_{z,av}^{''} < 0$ ($102<\omega_{ce}t <186$), and attains a quasi-steady value when $v_{z,av}^{''}$ becomes negligibly small. These observations suggest a mechanism different from viscous like dissipation  (which would require $v_{z,av}^{''} > 0$) is operating  at least after $v_{z,av}^{''}$ becomes negative or may be throughout the evolution. The scatter plot of the values of $|dv_{z,av}/dt|$ and $|v_{z,av}|$ in the close neighborhood of $x=0$ during the time period $102 \leq \omega_{ce}t \leq 186$ shows a linear correlation between the two quantities (see Fig. \ref{fig:jzav}d). The points deviating from the linear relationship are from the beginning and the end of the chosen time interval. Assuming resistive-like dissipation $dv_{z,av}/dt=-\nu_{c}v_{z,av}$, one can estimate from Fig. \ref{fig:jzav}d the anomalous collision frequency to be $\nu_{c}/\omega_{ce} \approx 0.05$. 
Altogether one can say  that the back reaction of the turbulence, caused by the instability, on the current profile manifests itself as both  viscous and resistive-type dissipation.

\begin{figure}
\includegraphics[width=0.7\textwidth,height=0.8\textheight]{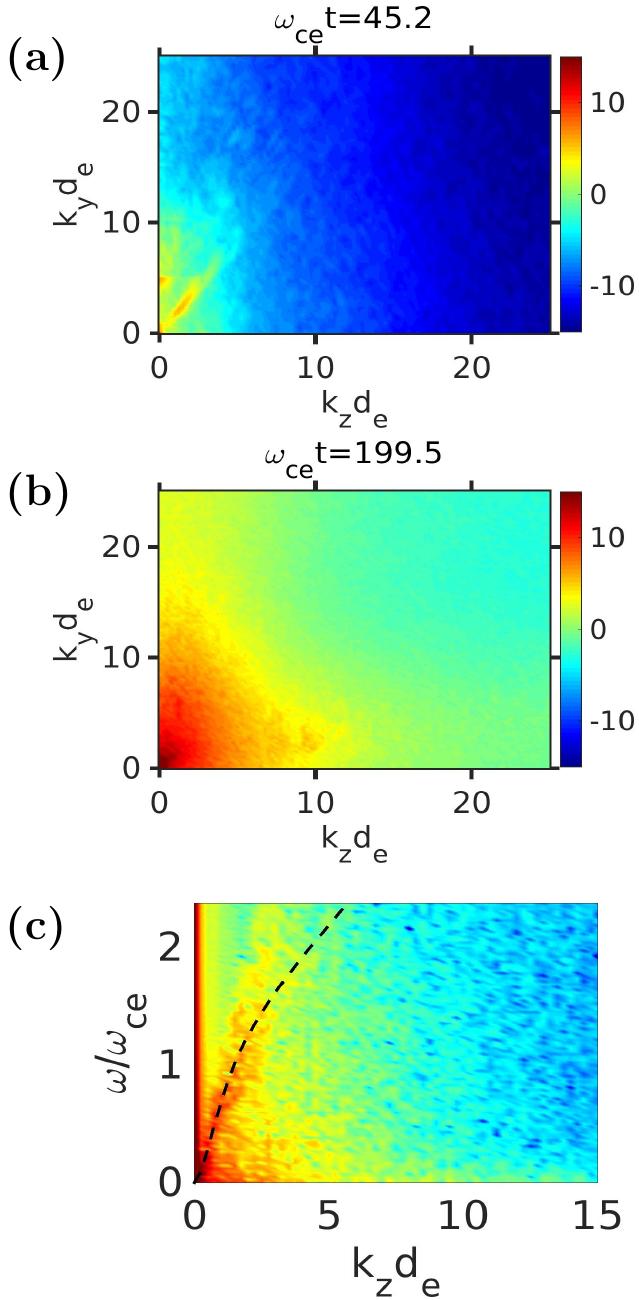}

\caption{\label{fig:spectra}   $k_x$-averaged power spectra of the magnetic field perturbations in the early (a) and late (fully nonlinear) (b) phase of the evolution for $L=d_e$ and $B_g=B_0$. The power of the magnetic field (color coded) in frequency-wave number ($\omega-k_z$) domain for $k_y=0$ (c). The dashed line in (c) shows the results of the local dispersion relation. 
}
\end{figure}

Figs. \ref{fig:spectra}a and \ref{fig:spectra}b depict the $k_x$-averaged power spectra of the perturbed magnetic field, $\bar{P}_{\tilde{B}}=\int \tilde{B}_k^2(k_x,k_y,k_z) \, dk_x$ where $\tilde{B}_{k}(k_x,k_y,k_z)$ is the Fourier transform of $\tilde{B}(x,y,z)$. It shows that in the early phase of the evolution (at $\omega_{ce}t \approx 45$), although the wave power is mostly concentrated around the linearly unstable growing modes, a nonlinear transfer of power to the linearly stable modes has already started. In the late nonlinear phase of the evolution (at $\omega_{ce}t\approx200$), the power distribution becomes anisotropic due to a preferential transfer of power to short wavelength perturbations in the y-direction (perpendicular to the guide magnetic field).

\begin{figure}[h]
\includegraphics[width=0.7\textwidth,height=0.8\textheight]{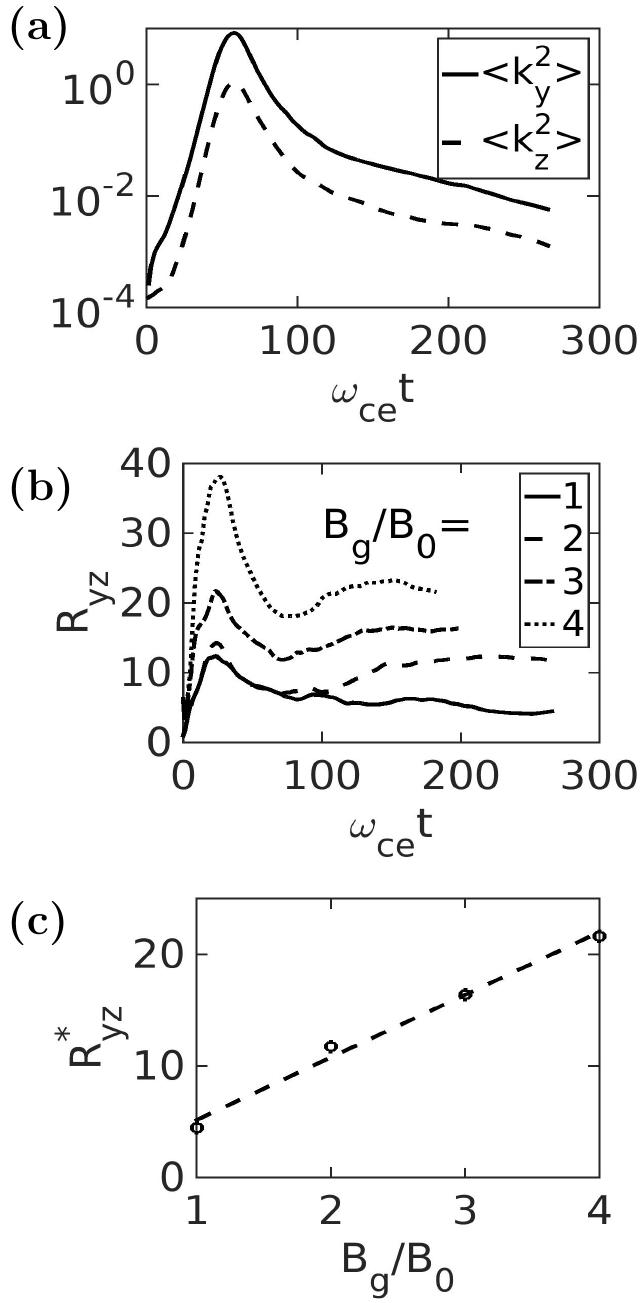}

\caption{\label{fig:anisotropy} Evolution of mean-square wave numbers $<k_y^2>$ and $<k_z^2>$ for thin ECS ($L=d_e$) and $B_g=B_0$ (a) and of the anisotropy ratio $R_{yz}=<k_y^2>/<k_z^2>$ for $L=d_e$ but different guide fields (b). The scaling of the anisotropy ratio with guide field strength (c).}
\end{figure}

Fig. \ref{fig:anisotropy}a and \ref{fig:anisotropy}b shows the temporal evolution of mean-square wave numbers defined as
\begin{eqnarray}
<k_{y,z}^2>=\frac{\int k_{y,z}^2\tilde{B}_k^2(k_x,k_y,k_z) dk_x\,dk_y\,dk_z}{\int \tilde{B}_k^2(k_x,k_y,k_z) dk_x\,dk_y\,dk_z},
\end{eqnarray}
and of the  anisotropy ratio $R_{yz}=<k_y^2>/<k_z^2>$.
Both $<k_y^2>$ and $<k_z^2>$ grow to attain  peak values and then fall until they finally saturate in the nonlinear phase of the instability evolution. The peaks in $<k_y^2>$ and $<k_z^2>$ are reached mainly due to the excitation of linearly unstable modes in the early phase of the evolution. The value of $<k_z^2>$ remains smaller than that of $<k_y^2>$ throughout the evolution indicating anisotropic distribution of power. The anisotropy ratio $R_{yz}$ in Fig. \ref{fig:anisotropy}b shows that $<k_y^2>$ remains greater than $<k_z^2>$ throughout the evolution for strength of guide field in the range $B_g/B_0=$1-4.    In the early phase this is because of $k_y > k_z$ typical for the linearly unstable modes in case of   finite guide fields \cite{jain2015}. This can be seen  in Fig. \ref{fig:spectra}a: the power in linearly unstable modes is confined to $k_y>k_z$.  In the nonlinear phase, the anisotropy $R_{yz} > 1$ is due to the interaction of whistler- and flow-induced-wave modes. This leads to a preferential transfer of the power to short scales in the direction perpendicular to the mean electron flow and the mean magnetic field direction \cite{dastgeer2000a,dastgeer2000b,jain2012b}. Fig. \ref{fig:anisotropy}c shows the saturated value of $R_{yz}$ vs. the strength of the guide field. The linear increase of saturated $R_{yz}$ with the guide field strength is in agreement with the earlier results of the  scaling of anisotropy ratio with the external magnetic field \cite{dastgeer2003}. 

In the electron shear flow instability  driven turbulence, flow induced new wave modes along with whistler wave modes can contribute to the anisotropic cascade. The new wave modes can be identified using the  EMHD local dispersion relation. The latter can be written as \cite{jain2015,jain04},
\begin{eqnarray}
\bar{\omega}&=&\frac
{k_z(d_e^2v''-v)\pm \sqrt{k_z^2(d_e^2v''-v)^2+4d_e^4\omega_{ce}^2(F''+k^2F)(F-d_e^2F'')/B_0^2}}
{2(1+k^2d_e^2)}
\label{eq:local_disp}
\end{eqnarray}
where $\bar{\omega}=\omega-k_zv$, $k^2=k_x^2+k_y^2+k_z^2$ and $F=\mathbf{k}.\mathbf{B}$. In the absence of the electron flow ($v=-B'$ and its higher derivatives vanish) Eq. (\ref{eq:local_disp}) becomes the well-known whistler-mode dispersion relation. In the limit $\mathbf{B}\rightarrow 0$, one obtains the dispersion relation of  the flow induced wave modes \cite{das01}. 
 \begin{eqnarray}
\omega&=&\frac
{k_zd_e^2(v''+k^2v)}
{1+k^2d_e^2}
\label{eq:flow_disp}
\end{eqnarray}
When $v\neq 0$ and $B\neq 0$, both the flow and the magnetic field contribute to the wave frequency. Fig. \ref{fig:spectra}c shows the (color-coded) dispersion relation obtained by the simulations. Over-plotted by a dashed line is the local dispersion relation, calculated according to Eq. \ref{eq:local_disp}, for $k_x=k_y=0$, $B_y=0$, $B_z=B_g$, $v=0.25$, $v'=0$ and $v''=0$. One can see that the simulated and the analytically obtained dispersion relations are in good agreement. Note that finite power at high frequencies ($\omega > \omega_{ce}$) is due to the flow-induced wave modes as the whistler frequency is bounded by the electron cyclotron frequency ($\omega_{whistler} < \omega_{ce}$).

Note that the anisotropic power cascade towards short wavelengths perpendicular to the z-direction is due to both the mean electron flow and the mean, external, guide magnetic field. In the present simulations, the electron flow and the mean magnetic field near the center of the sheet are both directed in the same z-direction. Thus the anisotropic power cascade perpendicular to the z-direction is mediated by  both, whistler and flow-induced waves together. Note that for ESFI in anti-parallel magnetic fields ($B_g=0$), the mean magnetic field at the center of the current sheet is too weak and thus the anisotropy is caused primarily by the flow induced wave modes \cite{jain2012b,jain2014b}.
\section{Summary\label{sec:conclusion}}
We investigated the nonlinear evolution of 3-D electron shear flow instabilities in the presence of a finite (external) guide magnetic field pointing in the direction of the electron current, out of the reconnection plane. Starting with the linear wave-mode analysis and carrying out 3-D EMHD simulations we found that a finite guide field subdues the tearing mode even for the thin current sheets with a half-thickness equal to an electron inertial length. Instead, in the presence of finite guide fields the current sheet evolution  is dominated by non-tearing wave mode instabilities which cause the formation of electron vortices and current filaments. This reminds the instabilities in case of thicker current sheets $L>d_e$ in anti-parallel magnetic fields (vanishing guide field). The resulting turbulence becomes more anisotropic with the larger  guide fields. It provides viscous (flattening and broadening  the current sheet) and resistive (localized reduction in the current density causing bifurcation) type dissipation.   

\begin{acknowledgments}
This work was supported by Max-Planck/Princeton Center for Plasma Physics at the Max Planck Institute for Solar System Research, Justus-von-Liebig-Weg-3, G\"ottingen, Germany and by the German Science Foundation CRC 963, project A03. 
\end{acknowledgments}

\appendix*

\bibliography{references}

\end{document}